\documentclass[prd,nofootinbib,showpacs,twocolumn,floatfix]{revtex4}
\usepackage{latexsym}
\usepackage{dcolumn}
\RequirePackage{graphicx}

\newcommand{\MET}{{\slash\!\!\!\!E_T}}
\newcommand{\DZero}{D$\slash\!\!\!0$}
\newcommand{\fbi}{\mathrm{fb}^{-1}}

\newcommand{\be}{\begin{equation}}
\newcommand{\ee}{\end{equation}}
\newcommand{\bea}{\begin{eqnarray}}
\newcommand{\eea}{\end{eqnarray}}

\begin{document}

\preprint{IIT-CAPP-11-09}

\title{Higgs exclusion and the $H \to WW^*\to l\nu c j$ semi-leptonic
  channel at the Tevatron}

\author{Arjun~Menon}
\affiliation{Illinois Institute of Technology,
Chicago, Illinois 60616-3793, USA}
\affiliation{Department of Physics, University of Oregon, Eugene, OR 97403,USA}
\author{Zack~Sullivan}
\email{Zack.Sullivan@IIT.edu}
\affiliation{Illinois Institute of Technology,
Chicago, Illinois 60616-3793, USA}

\date{September 29, 2011}

\begin{abstract}
  We study the Higgs boson decay to $W^+W^-$, where one boson decays
  to leptons, and the other decays to $c+$jet at Tevatron.  Given the
  current charm tagging acceptances, this channel can help improve and
  confirm the current combined Tevatron exclusion limit on a standard
  model-like Higgs boson.  If charm acceptance can be improved to at
  least 24\%, this channel could provide the second tightest limits on
  a Higgs boson mass between 140--190 GeV from a single channel
  measurement.
\end{abstract}


\maketitle

\section{Introduction}
\label{sec:introduction}

As data collection concludes at the Fermilab Tevatron, the CDF and
\DZero\ experiments are combining their results to place strong
constraints on the existence of a standard model-like Higgs boson
\cite{CDF:2011cb}.  The strongest constraints come from the search for
$H\to WW\to l^+l^-\nu\bar\nu$
\cite{Barger:1990mn,Dittmar:1996ss,Han:1998ma}, setting preliminary
Tevatron mass exclusion limit of 156--177 GeV
\cite{CDF:pub11,DZERO:pub11,CDF:2011cb}.  Several other modes have
also been measured in order to contribute to this limit
\cite{CDF:pub10573,CDF:pub10574}, but have significantly less reach.
In this paper we propose that the Tevatron experiments add the channel
$H \to l \nu c j$ to the final combination limit.  We first proposed
in Ref.\ \cite{Menon:2010vm} $H\to l\nu cj$ for analysis at the CERN
Large Hadron Collider (LHC) as a useful addition to the Higgs search,
as well as a motivation to improve charm tagging.  In this paper, we
demonstrate this channel already has comparable reach at the Tevatron
to several of the other channels measured for the combined analysis,
and, with a little work, could be the second most powerful channel in
the 140--190 GeV region.

While impressive limits on the Higgs mass have been set using the
$H\to WW\to l^+l^-\nu\bar\nu$ channel, a search for the Higgs in the
$H\to l\nu cj$ final state has a number of important attributes.
Foremost, the $H\to l\nu cj$ final state can be fully reconstructed to
provide a Higgs mass peak.  The dilepton plus missing energy $\MET$
search relies on fitting derivative shapes, such at the transverse
mass of the dilepton pair $M_T^{ll}$.  Second, while the absolute size
of the backgrounds is higher, we show in Sec. \ref{sec:results} the
signal to background ratio (S/B) for $H\to l\nu cj$ is a factor of
2--4 better than for the $H\to l^+l^-\MET$ case.  Third, $H\to l\nu cj$ is
sensitive to an independent set of backgrounds, providing a robustness
check on the Higgs mass limits.  This may be important, as recent
loosening of cuts in the existing Tevatron analyses \cite{CDF:2011cb}
may have reopened a sensitivity to the background due to heavy-quark
decays into isolated leptons \cite{Sullivan:2006hb}.

\begin{figure}[htb]
\centering
\includegraphics[width=0.4\columnwidth]{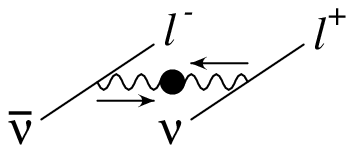}
\includegraphics[width=0.4\columnwidth]{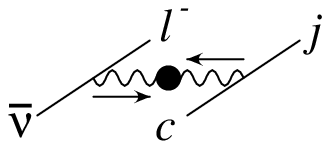}
\caption{Angular correlation between leptons in (left) $H\to W^-W^+\to
  l^-\bar\nu l^+\nu$ and (right) $H\to W^-W^+\to l^-\bar \nu c j$ due
  to the anti-alignment of $W$ boson spins in Higgs boson decay.
\label{fig:WW}}
\end{figure}

A final advantage of the $H\to l\nu cj$ channel over the
dilepton$+\MET$ channel is the ability to make use of additional
angular correlations in the final state.  The key ingredient to
reducing both $WW$ \cite{Barger:1990mn,Dittmar:1996ss,Han:1998ma} and
QCD \cite{Sullivan:2006hb} backgrounds in dilepton$+\MET$ comes about
because in a spin-0 Higgs boson decay to $WW$, the spins of the $W$
bosons are anti-aligned.  The pure $V-A$ structure of $W$ boson decay
causes a strong enhancement of the cross section when the charged
leptons are aligned.  There is also an enhanced probability that the
neutrinos align, as seen in Fig.~\ref{fig:WW}, but this information is
lost as the neutrinos are unobservable.

In Ref.~\cite{Menon:2010vm} we demonstrated that charm tagging can be
used to uniquely identify the correlations between all particles in
the final state.  In Fig.~\ref{fig:WW} we see one of the neutrinos of
the $H\to l^+l^-\nu\nu$ channel is replaced by a charm quark.  This allows
us to make use of both a strong correlation between the charged lepton
and the light-quark jet, and between the charmed jet and the neutrino
in the event.  Despite the fact that the neutrino appears as missing
energy, we examine cases where it comes from an on-shell $W$ decay,
and can reconstruct its four-momentum up to a two-fold ambiguity in
rapidity.

We have motivated the $H\to l\nu cj$ channel as complementary to
dileptons$+\MET$.  In Sec.~\ref{sec:sim} we focus on our simulation of
the $l\nu cj$ signal and backgrounds.  There we stress the equivalence of
our both our modeling of charm tagging and jet energy corrections to
existing Tevatron algorithms.  We also describe a parametrization of
the tagging efficiencies and fake rates we use to predict what benefits
could be derived from an increase in charm acceptance.  In
Sec.~\ref{sec:results} we optimize our cuts in three regions,
corresponding to whether the $W$ bosons are off-shell, nearly
on-shell, or on-shell, and present the Higgs mass reach.  We conclude
in Sec.~\ref{sec:concl} by placing our predictions for Higgs mass
reach as a function of different charm tagging efficiencies in the
context of existing measurements from the Fermilab Tevatron.

\section{Jet energy scale corrections, charm tagging, and other
  simulation details}
\label{sec:sim}

Our simulation of signals and backgrounds closely follows our analysis
of $H\to l\nu cj$ at the LHC \cite{Menon:2010vm}, with a few
improvements for jet energy scales.  In order to retain all angular
correlations, we generate events at a $\sqrt{s}=1.96$~TeV $p\bar p$
collider using MadEvent 4.4~\cite{Alwall:2007st} and CTEQ6L1 parton
distribution functions (PDFs) \cite{Pumplin:2002vw}.  We shower the
events with PYTHIA 6.4~\cite{PYTHIA} and use the PGS 4~\cite{PGS4}
detector simulation to reconstruct leptons and jets.  Angular
correlations in the Higgs signal tend to force the lepton and leading
non-tagged jet to be close in phase space.  Hence, we reconstruct jets
using the PGS jet cone algorithm with a cone size of $0.4$.  Missing
transverse energy $\MET$ is reconstructed from the calorimeter and
corrected for muons.  Charm tagging efficiencies and fake rates in PGS
are replaced with the ones described below.

In accordance with the experimental analyses, soft jets reconstructed
by PGS require large jet energy scale (JES) corrections.  The same
angular corrections that lead to soft leptons in $H\to l^+l^-\MET$
studies, lead to soft jets in our study.  Hence, to correct for
underestimated gauge and Higgs boson masses in our analysis we
calculate a jet energy scale correction as a function of the
transverse energy $E_T$ of the jet.

\begin{figure}[htb]
\centering
\includegraphics[width=3in]{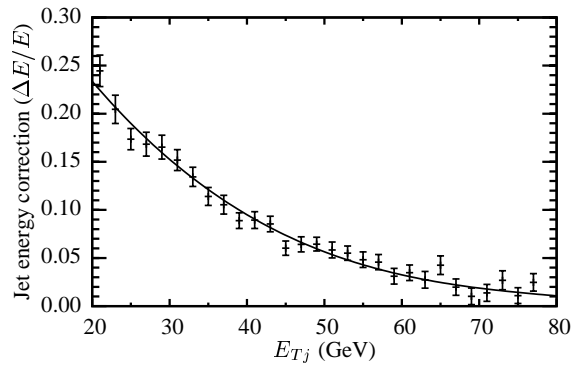}
\caption{Jet energy scale correction as a function of the jet transverse
energy $E_{Tj}$.}
\label{fig:jec}
\end{figure}

We derive the JES correction by fitting the shift in the energy
between the leading parton generated by $Zj\to e^+e^-j$ in MadEvent
and the leading jet reconstructed by PGS using a cone size of $0.4$.
In Fig.~\ref{fig:jec} we show the data for this extracted jet energy
correction with statistical errors, and the best fit curve to this
data.  The correction is large for low energy jets, and is numerically
almost identical to the CDF jet energy correction
\cite{Bhatti:2005ai}.  We confirm that by applying our JES
correction, the Higgs, $W$, and $Z$ gauge boson mass peaks are
reconstructed to their correct central values.  After we apply the JES
correction to the raw PGS output, the following acceptance cuts are
used to define jets and leptons: \be E_T^j > 20~\mathrm{GeV}, \;
|\eta_j| < 2.0; \; p_T^l > 20~\mathrm{GeV}, \; |\eta_l| < 2.0 \;.  \ee

In contrast to the main dilepton$+\MET$ studies, we allow an
additional jet to be in the event --- consistent with the effects of
next-to-leading order (NLO) radiation.  We normalize our cross
sections to the NLO cross sections obtained after acceptance cuts
applied in MCFM 5.8~\cite{MCFM} using CTEQ 6.5 PDFs
\cite{Tung:2006tb}.  Effective $K$-factors after cuts are shown in
Table~\ref{tab:kfac}.

\begin{table}[htb]
\caption{Next-to-leading order $K$-factors (after acceptance cuts) for
the signal and backgrounds.\label{tab:kfac}}
\begin{ruledtabular}
\begin{tabular}{ccccccccc}
Signal & $Wcj$ & $WW$ & $t\bar{t}$ & $Wbj$ & t(s)-chan.& $Wc\bar{c}$ & 
$Wb\bar{b}$ & $Wjj$\\
&&&&& single top &&&\\
\hline
2.19 & 1.39 & 1.32 & 1.2 & 1.59 & 0.96(1.52) & 1.59 & 1.59 & 1.39
\end{tabular}
\end{ruledtabular}
\end{table}

The key to the measurement of the Higgs boson in the $H\to l\nu cj$
channel is charm tagging.  More specifically, the key is charm
\textit{acceptance}, as the dominant background will turn out to be
direct $Wcj$ production.  Hence, we explore several possible tagging
and fake rate efficiencies in order to map out the possible spectrum of
results.

We model the transverse energy $E_T$ dependence of the tagging
efficiencies utilizing existing impact parameter $b$-tagging
algorithms based on Tevatron Run I codes \cite{Carena:2000yx} (as
appear in PGS 3.2 and earlier), and Run II (as appear in PGS 4)
\cite{PGS4}.  Our main results use rescaled Run I-like charm and
bottom tagging efficiencies of the form
\bea
\epsilon^1_c &=&  k_c \times 0.2 \tanh \left(
\frac{E_{Tj}}{42.08 \mathrm{\ GeV}}\right) \;, \nonumber\\
\epsilon_b &=& \min \left[ 1.0, k_b \times 0.6\tanh\left(
\frac{E_{Tb}}{36.05 \mathrm{\ GeV}}\right) \right] \;,
\eea
where $k_c=1$, $k_b=1$ correspond to current tagging efficiencies.
This heavy-flavor tagging algorithm is predominantly a fit to
distributions of events in impact parameter vs.\ track invariant mass
\cite{wicklund}, and has room for improvements to acceptance by varying
the reconstruction cuts.  We scale $k_b = (k_c+3)/4$ to model overall
increases in heavy flavor acceptance.  Eventually, 100\% of $b$ jets
will be retained as background, but this is an advantage, as we will
want to veto events with two heavy-flavor tags.  We also assume a
constant light jet fake rate $\epsilon_j = 1\% \times 10^{(k_c-4)/5}$.
The default choices here are consistent with the CDF Run II measurement
of $Wc$ \cite{:2007dm}, whose kinematics are similar to ours.

\begin{figure}[htb]
\centering
\includegraphics[width=3in]{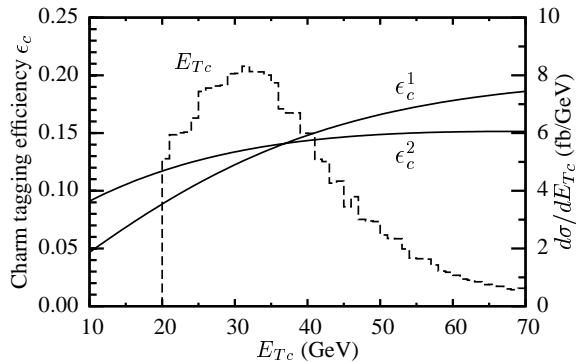}
\caption{Charm tagging efficiency curves and characteristic charm jet energy, 
$E_{Tc}$,
from $H\to WW\to l\nu c j$ decay. $\epsilon^1_c$($\epsilon^2_c$) are existing
Tevatron Run I(II)-like algorithms.
\label{fig:ctag}}
\end{figure}

As in Ref.\ \cite{Menon:2010vm}, we have reproduced all results using
a Run II-like algorithm $\epsilon^2_c$ from PGS 4.  The $E_T$
dependence of the charm tagging efficiencies for $\epsilon^1_c$ and
$\epsilon^2_c$ are shown in Fig.\ \ref{fig:ctag}.  Despite the
different $E_T$ dependence, we find exactly the same significances
after cuts.  Over the kinematic range of the signal charm jets, the
charm acceptance is currently $\sim 12\%$ for both algorithms; and so
we present the full details using $\epsilon^1_c$.  It is likely that
modern neural network based tagging algorithms could improve both the
acceptance and purity of the charm signal; but we stress again, that
the acceptance is the key to this analysis.  In Sec.\
\ref{sec:results} we provide a complete list of backgrounds, and our
predictions can be trivially rescaled to the final results.

\section{Analysis and mass reach for $H\to l\nu cj$ at the Tevatron}
\label{sec:results}

The initial sample for this analysis contains exactly one isolated
lepton (electron or muon), two or three jets with exactly one charm
tag, and missing transverse energy $\MET > 15$~GeV.  The standard
model backgrounds for the $lcj+\MET$ final state include $Wcj$, $Wbj$,
$Wjj$, $Wc\bar{c}$, $Wb\bar{b}$, $W^+ W^-$, $t$- and $s$-channel
single top, and $t\bar{t}$.  By restricting the number of jets to two
or three, we reduce significantly the $t\bar{t}$ background.  Charm
tagging substantially reduces the $Wjj$ background, leaving $Wcj$ as
the most important background at most levels of cuts.

\begin{figure}[htb]
\centering
\includegraphics[width=3in]{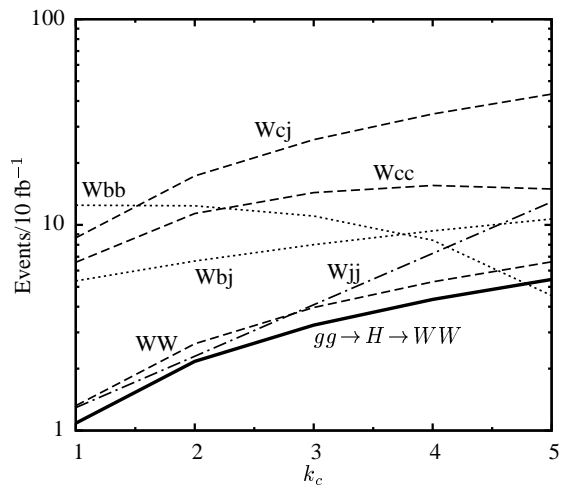}
\caption{Number of events in 10 $\fbi$  at a $1.96$ TeV Tevatron for the 
$H\to l\nu cj$ signal (solid), and backgrounds from processes with charm jets
(dashed), bottom jets (dotted), and light jets (dot-dashed).
\label{fig:modes}}
\end{figure}

Unlike at the LHC, where the $Wcj$ is the overwhelming dominant
background for any charm acceptance, the $Wbb$, $Wbj$, and $Wcc$
backgrounds are important at the current low charm-tagging efficiency.
Hence, $S/B$ of this channel at the Tevatron in is not as independent
of the charm tagging efficiency as that at the LHC.  This is evident
in Fig.\ \ref{fig:modes}, where we show the number of events expected
for the signal and each background as a function of $k_c$.
Nevertheless, the direct $Wcj$ background scales with the signal, and
so our analysis concentrates mostly on reducing its effects.

We optimize our cuts in the three different Higgs mass regions $m_H <
160$~GeV, $160 \leq m_H < 170$~GeV, and $m_H \geq 170$~GeV.  For each
of these regions the sequence of cuts is similar, however the
strengths of the cuts are slightly different so as to emphasize the
optimal reach at $m_H = 150$~GeV and $m_H = 180$~GeV --- outside the
current Tevatron limits.  To compare the Tevatron reach in the $H \to
lcj+ \MET$ channel with the dilepton and other measured channels, we
also optimize a search for $m_H = 165 $~GeV.

As $Wcj$ is the most problematic background, we tune most cuts to
reduce its contribution.  In the region $m_H < 160$~GeV we optimize
our cuts for $m_H = 150$~GeV, which are shown in Tab.~\ref{tab:mh150}.
To be concrete, all numbers are presented for $k_c=4$, but we present
results for all charm tagging efficiencies below.  In this region, at
least one of the $W$ gauge bosons is off-shell, and the signal is
typically softer than in the higher mass regions.  Therefore, we can
impose tight cuts on the upper bounds of $\MET$, the charm jet
transverse energy $E_{Tc}$, and lepton transverse momentum $p_{Tl}$.
These three cuts reduce the $Wjj$ oriented backgrounds by a factor of
3, and nearly eliminate $t\bar t$.

\begingroup
\squeezetable
\begin{table*}[htb]
  \caption{Number of signal ($H\to l\nu cj$) and background events per
    10~fb$^{-1}$ of data for $m_H=150$~GeV with $k_c=4$.\label{tab:mh150}}
\begin{ruledtabular}
\begin{tabular}{l|ddddddddd}
Cuts & \text{Signal} & Wcj & WW & t\bar{t} & Wbj & \text{Single top} & 
Wc\bar{c} & Wb\bar{b} & Wjj\\
\hline
1 $l$, 2 or 3 jets w/ 1 $c$ tag & 12 & 1434 & 162 & 31 & 252 & 461 & 442 & 
334 & 816\\
$15 \mathrm{\ GeV} < \MET < 50$~GeV & 11 & 1134 & 104 & 8.5 & 189 & 258 & 314 & 
235 & 464\\
$E_{Tc} < 65$~GeV & 9.6 & 716 & 67 & 3.5 & 132 & 113 & 209 & 160 & 150 \\
$p_{Tl} < 60$~GeV & 8.0 & 354 & 39 & 1.8 & 75 & 57 & 118 & 88 & 76\\
$\cos \theta_{jl}  > -0.2$ & 5.9 & 185 & 19 & 0.9 & 44 & 33 & 67 & 49 & 43\\
$\cos \theta_{c\nu} > -0.8$ & 5.7 & 172 & 17 & 0.8 & 40 & 30 & 60 & 44 & 40 \\
$M_{jc} < 100$~GeV & 5.4 & 148 & 17 & 0.5 & 33 & 20 & 52 & 40 & 32\\
$120 \mathrm{\ GeV} \leq M_{l\nu cj} \leq 200 \mathrm{\ GeV}$ & 5.3 & 144 & 17 & 0.5 & 32 & 19 & 50 & 39 & 31 
\end{tabular}
\end{ruledtabular}
\end{table*}
\endgroup

The backgrounds that are independent of charm tagging --- $Wb\bar{b}$,
$Wbj$, $t\bar t$, and single top --- can be reduced significantly by
using the angular correlations between the final state particles.  The
simplest angular cuts, are those similar to that of $\Delta \phi_{ll}$
in the leptonic channel, where we cut on $\cos \theta_{jl}$ the angle
between the lepton $l$ and \textit{leading} non-tagged jet $j$ in the
lab frame.  In addition, we know that the directions of the neutrino
and $c$-jet are correlated.  Therefore we can also make a cut $\cos
\theta_{c\nu}$.  In the low Higgs mass region, off-shell $W$ bosons
have weak angular correlations between the jets and leptons.  Hence,
we make weaker cuts on these angles and on the invariant mass of the
jet charm system than for heavier Higgs masses.

In order to reconstruct a Higgs mass peak, we reconstruct the neutrino
four-momentum $p_\nu$ by fitting the lepton and $\MET$ to an on-shell
$W$ boson mass.  We take the smallest absolute rapidity $|\eta_\nu|$
solution to complete the fit.  We finish the low-mass Higgs search by
placing a cut on the $M_{l\nu cj}$ invariant mass.

In the region $m_H \geq 160$~GeV both $W$ gauge bosons are on-shell.
This condition lends itself to slightly different optimizations, as
the objects in the final state have more energy on average.  In the
region $160 \leq m_H < 170$~GeV, shown in Tab.~\ref{tab:mh165}, we
loosen the upper cut on the $\MET$.  However, the on-shell condition
strengthens the angular correlations.  We use this to tighten cuts on 
$\cos \theta_{jl}$ and $\cos \theta_{c\nu}$.

\begingroup
\squeezetable
\begin{table*}[htb]
  \caption{Number of signal ($H\to l\nu cj$) and background events per
    10~fb$^{-1}$ of data for $m_H=165$~GeV with $k_c=4$.\label{tab:mh165}}
\begin{ruledtabular}
\begin{tabular}{l|ddddddddd}
Cuts & \text{Signal} & Wcj & WW & t\bar{t} & Wbj & \text{Single top} & 
Wc\bar{c} & Wb\bar{b} & Wjj\\
\hline
1 $l$, 2 or 3 jets w/ 1 $c$ tag & 16 & 1434 & 162 & 31 & 252 & 461 & 442 & 
334 & 816\\
$15 \mathrm{\ GeV} < \MET < 55$~GeV & 15 & 1209 & 110 & 11 & 193 & 330 & 
579 & 218 & 520 \\
$E_{Tc} < 65$~GeV & 13 & 721 & 84 & 4.9 & 144 & 165 & 410 & 160 & 157\\
$p_{Tl} < 60$~GeV & 11 & 361 & 51 & 2.5 & 83 & 92 & 232 & 86 & 78\\
$\cos \theta_{jl} > -0.3$ & 6.1 & 108 & 14 & 0.8 & 30 & 32 & 72 & 29 & 25\\
$\cos \theta_{c\nu} > -0.6$ & 5.6 & 86 & 10 & 0.5 & 23 & 23 & 49 & 22 & 18\\
$\cos \theta_{jc} < 0.8$ & 5.2 & 68 & 8.2 & 0.3 & 19 & 14 & 41 & 19 & 14\\
$50 < M_{jc} < 100$~GeV & 4.3 & 35 & 5.3 & 0.1 & 9.4 & 5.6 & 16 & 8.4 & 
7.3 \\
$135 \mathrm{\ GeV} \leq M_{l\nu cj} \leq 195 \mathrm{\ GeV}$ & 4.1 & 33 & 5.3 & 8.6 & 4.4 & 0.9 & 14 & 8.2 & 6.9
\end{tabular}
\end{ruledtabular}
\end{table*}
\endgroup

In addition to the low-mass cuts used in Tab.~\ref{tab:mh150}, once we
are above $WW$ threshold, the \textit{leading} non-tagged jet $j$ and
the $c$-jet have a large opening angle, which allows us to impose a
cut on $\cos \theta_{jc}$.  We also impose a weak $W$ mass
reconstruction cut on $M_{jc}$.  After Higgs mass reconstruction, the
signal to background ratio $S/B\sim 1/18$ for the 165~GeV Higgs boson
of in Tab.~\ref{tab:mh165}, where $k_c=4$.  The $S/B$ is $1/40$ with
current charm acceptance ($k_c=1$), and is a factor of 2 better than
the 0-jet dilepton channel \cite{CDF:pub11}.

For masses above 170~GeV, we optimize the cuts for $m_H = 180$~GeV.
As the objects in the events become harder, we loosen the upper limit
on $\MET$, $E_{Tc}$, and $p_{Tl}$.  However, we again compensate by
tightening the cuts on the angular correlations.  The results are
shown in Tab.~\ref{tab:mh180}.

\begingroup
\squeezetable
\begin{table*}[htb]
  \caption{Number of signal ($H\to l\nu cj$) and background events per
    10~fb$^{-1}$ of data for $m_H=180$~GeV with $k_c=4$.\label{tab:mh180}}
\begin{ruledtabular}
\begin{tabular}{l|ddddddddd}
Cuts & \text{Signal} & Wcj & WW & t\bar{t} & Wbj & \text{Single top} & 
Wc\bar{c} & Wb\bar{b} & Wjj\\
\hline
1 $l$, 2 or 3 jets w/ 1 $c$ tag & 14 & 1434 & 162 & 31 & 252 & 461 & 442 & 
334 & 816\\
$15 \mathrm{\ GeV} < \MET < 70$~GeV & 14 & 1351 & 140 & 15 & 235 & 408 & 399 & 
298 & 638 \\
$E_{Tc} < 80$~GeV & 12 & 969 & 105 & 8.9 & 191 & 261 & 310 & 234 & 298\\
$p_{Tl} < 75$~GeV & 11 & 710 & 81 & 6.1 & 139 & 203 & 232 & 173 & 204\\
$\cos \theta_{jl} > -0.3$ & 8.4 & 410 & 45 & 3.5 & 92 & 143 & 142 & 107 & 125\\
$\cos \theta_{c\nu} > -0.4$ & 7.2 & 296 & 28 & 2.4 & 63 & 96 & 92 & 71 & 77\\
$\cos \theta_{jc} < 0.8$ & 6.8 & 261 & 24 & 2.2 & 58 & 86 & 67 & 54 & 68 \\
$50 < M_{jc} < 100$~GeV & 5.8 & 158 & 21 & 1.0 & 35 & 40 & 34 & 30 & 39 \\
$150 \mathrm{\ GeV} \leq M_{l\nu cj} \leq 210 \mathrm{\ GeV}$ 
& 4.9 & 118 & 16 & 0.6 & 25 & 28 & 23 & 22 & 29
\end{tabular}
\end{ruledtabular}
\end{table*}
\endgroup

\begin{figure}[htb]
\centering
\includegraphics[width=3in]{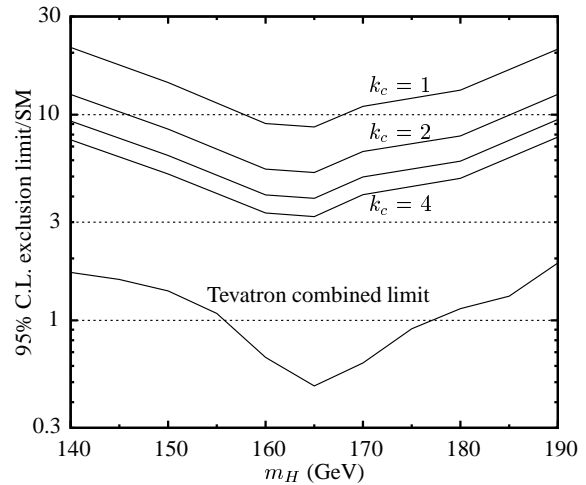}
\caption{Expected 95\% confidence level (C.L.) exclusion limit on the ratio
of the cross-section to that of the standard model (SM) for various charm 
tagging efficiencies vs.\ Higgs mass.  Also shown is the preliminary combined 
CDF and \DZero\ analysis including all channels at the Tevatron
\protect{\cite{CDF:2011cb}}.
\label{fig:reach}}
\end{figure}

In Tabs.~\ref{tab:mh150}--\ref{tab:mh180} we present the detailed
effects of cuts level-by-level for 150, 165, and 180~GeV Higgs bosons,
and assuming one experiment that collects 10~$\fbi$ of integrated
luminosity.  The CDF and \DZero\ experiments have made a strong effort
to combine their Higgs mass-limit analyses in order to improve their
reach.  In Fig.~\ref{fig:reach}, we scan masses from 140--190 GeV, and
compare the 95\% exclusion reach of the $H\to WW\to l\nu cj$ channel
using 8.6~$\fbi$ per experiment to the current preliminary combined
Tevatron limits \cite{CDF:2011cb}.  We demonstrate the importance of
the charm acceptance, by presenting separate curves for current charm
tagging acceptance $k_c=1$ through a best case acceptance of $\sim
48\%$ ($k_c=4$).  This channel alone can already reach $8.7$ times the
standard model cross section at 165~GeV current charm tagging, and
could reach $3.2$ times the standard model cross section if $k_c=4$.

\section{Conclusions}
\label{sec:concl}

A search for the Higgs boson in the channel $H\to l\nu cj$ could
provide significant additional information to the Fermilab Tevatron
Higgs mass exclusion limits.  While this channel cannot compete with
dileptons$+\MET$ in absolute rate, it has several strengths in
combination with other measurements.  The strongest features are that
the Higgs mass is completely reconstructable, and the signal to
background $S/B$ is roughly a factor of 2--3.5 better than
dileptons$+\MET$ \cite{CDF:pub11}.

\begin{figure}[htb]
\centering
\includegraphics[width=3in]{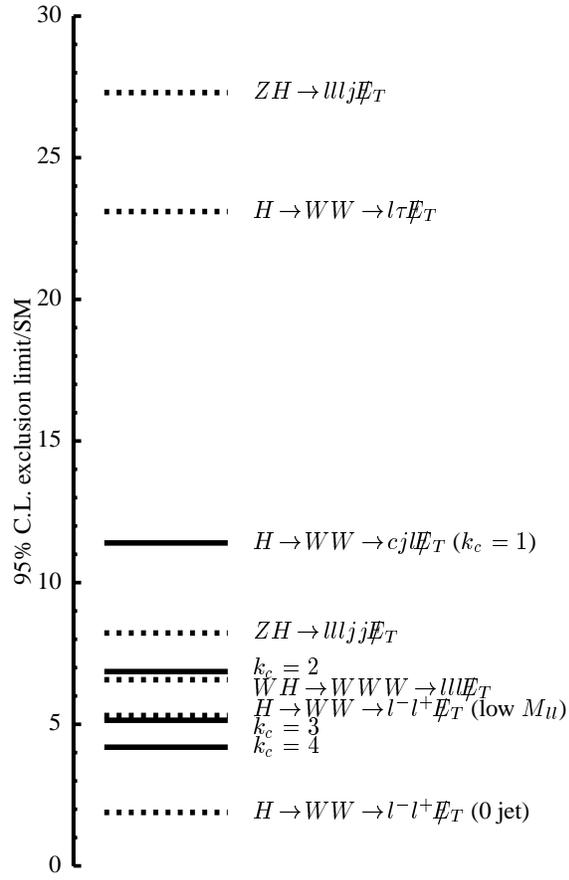}
\caption{Comparison of Higgs mass limit reach in $H\to cjl\MET$ to 
limits already extracted at the Fermilab Tevatron, for $m_H=165$ GeV.
\label{fig:vsother}}
\end{figure}

To emphasize the complementary nature of this channel, we compare
$H\to l\nu cj$ to several other channels already studied at the
Tevatron.  While $H\to lcj\MET$ is only sensitive to about 9 times
the standard model cross section at current charm acceptances, this is
better than several channels in the combined fits.  Recent measurements
of $H\to ZZ\to l^+l^-l^+l^-$ and $ttH$ are sensitive only to cross sections
40--60 times the standard model cross section
\cite{CDF:pub10573,CDF:pub10574}.  In Fig.\ \ref{fig:vsother}, we show
the 95\% confidence level exclusion limits from several measurements
\cite{CDF:2011cb,DZERO:pub11}.  If charm tagging acceptance could be
improved a factor of 2 ($\epsilon_c \sim 24\%$), $H\to lcj\MET$ would
have comparable reach to $WH\to WWW$, currently the second most
powerful individual channel.

As data taking at the Fermilab Tevatron comes to a close, the CDF and
\DZero\ Collaborations will work to produce strong combined limits on
a standard model-like Higgs boson mass.  For these final searches, we
recommend adding the channel $H\to l\nu cj$ to the list of
measurements used in the combinations.  Not only is this channel
interesting by itself, but its sensitivity to completely independent
backgrounds will enhance the robustness of the limits.

\begin{acknowledgments}
  This work is supported by the U.~S.\ Department of Energy under
  Contract Nos.\ DE-FG02-94ER40840 and DE-FG02-96ER40969.
\end{acknowledgments}

\end{document}